\documentclass[reprint,aps,prb]{revtex4-2}
\usepackage{bm,physics,amsmath,amssymb,
	graphicx,xcolor,embedfile,comment}
\usepackage[utf8]{inputenc}
\usepackage[normalem]{ulem}
\embedfile{main.tex}

\graphicspath{{figures/}}

\def\sout{\bgroup\markoverwith
{\textcolor{red}{\rule[0.5ex]{2pt}{0.5pt}}}\ULon}

\begin{document}

\title{Impurity with a resonance in the vicinity of the Fermi energy}

\author{Mikhail Maslov}
\email{mikhail.maslov@ist.ac.at}
\author{Mikhail Lemeshko}
\email{mikhail.lemeshko@ist.ac.at}
\author{Artem G. Volosniev}
\email{artem.volosniev@ist.ac.at}
\affiliation{IST Austria (Institute of Science and Technology Austria),
    Am Campus 1, 3400 Klosterneuburg, Austria}
    
\begin{abstract}
We study an impurity with a resonance level whose position coincides with the Fermi energy of the surrounding Fermi gas. An impurity causes a rapid variation of the scattering phase shift for fermions at the Fermi surface, introducing a new characteristic length scale into the problem. We investigate manifestations of this length scale in the self-energy of the impurity and in the density of the bath. Our calculations reveal a model-independent deformation of the density of the Fermi gas, which is determined by the width of the resonance. To provide a broader picture, we investigate time evolution of the density in quench dynamics, and study the behavior of the system at finite temperatures. Finally, we briefly discuss implications of our findings for the Fermi-polaron problem.
\end{abstract}

\maketitle

\section{Introduction}

Systems with defects and impurities are ubiquitous in modern condensed matter physics~\cite{Coleman,Mahan,Altland_Simons}. Already simple impurity models such as a potential interacting with a continuum of states have led to a number of important discoveries, such as Friedel oscillations~\cite{Friedel}. These simple models usually have a single length scale,~e.g.,~the range of the potential, and are not able to describe an impurity whose structure introduces other relevant length scales. For example, they fail to provide insight into the physics of the Kondo model~\cite{Kondo_review} (and the Anderson model~\cite{Anderson_impurity}), which enjoys an emergent length scale defined by the Kondo temperature. Such additional length scales are often crucial for understanding the physics of realistic systems. 

A basic impurity model with a non-trivial length scale is the Friedel impurity with the resonance at the Fermi level. 
In spite of its simplicity, numerical analysis shows that real-space properties of that model have some similarities to those of the Kondo model~\cite{Bergmann2008,Tao2012} whose understanding represents a milestone in condensed matter physics~\cite{Affleck2010,Borzenets2020}. 
Motivated by these numerical findings, we discuss here an analytically tractable model of an impurity that also features a resonance at the Fermi level. We discuss how the resonance affects properties of the Fermi gas, in particular the self-energy of the impurity, and the density profile of the Fermi gas. Our results provide an illustration of how an impurity can introduce a length scale, which governs the physics at intermediate distances.

We formulate the problem in such a way that it can be engineered in cold-atom set-ups. In particular, we assume a continuous homogeneous Fermi gas with the impurity that can either be an external potential or a tightly trapped atom. In the latter case, the resonance can be a Feshbach resonance~\cite{Feshbach_review} whose width is much smaller than the Fermi energy. For instance, the typical Fermi temperature in experiments with ultracold~$^{6}$Li~atoms is about $1\;\mu$K (see, e.g.,~Ref.~\cite{Grimm2007}, and references therein), which means that narrow resonances whose width is about $0.2$~G~\cite{Strecker2003,Schunck2005} satisfy the condition.

The paper is organized as follows. In Sec.~\ref{sec_resonance}, we introduce a simple model of an impurity with a Breit-Wigner resonance~\cite{Breit_Wigner} (see also Ref.~\cite{Fano1961}), which allows us to formulate the problem that can be solved analytically. In Sec.~\ref{sec_energy}, we calculate the energy of the system using Fumi's theorem~\cite{Fumi}, and  discuss the energy scale associated with immersion of the impurity. In Sec.~\ref{sec_density}, we study the density of the fermionic bath, and show that it is strongly affected by the resonance. In particular, we observe that the density oscillations are influenced by an emergent length scale, which is controlled by properties of the resonance. To link our studies to current cold-atom experiments, we study time evolution of the density in a numerical quench experiment in Sec.~\ref{sec_QS}, and compute finite-temperature properties of the system in Sec.~\ref{sec_temp}. In Sec.~\ref{sec_MF}, we briefly explain implications of our findings for the Fermi-polaron problem by employing a mean-field-like approximation (see also the Appendix). In Sec.~\ref{sec_conclusion}, we summarize our findings and give an outlook.

\section{Model of an Impurity with a Resonance}
\label{sec_resonance}

We consider a heavy impurity immersed in an ideal Fermi gas. The Hamiltonian of the system in the coordinate representation reads as
\begin{equation}
\hat{\mathcal{H}}=-\frac{\hbar^2}{2m}\sum\limits_{i=1}^{N}\Delta_i+\sum\limits_{i=1}^{N}U(\bm{r}_i)\,,
\label{eq_hamiltonian_heavy}
\end{equation}
where $\Delta_i$ and $\bm{r}_i$ denote, respectively, the Laplacian and the coordinate of the $i$th fermion, $m$ is the mass of a fermion, $N$ is the number of particles in the bath, and $U$ is the impurity-fermion interaction potential.  Without loss of generality, we assume that the system is confined to a three-dimensional box of size $R$ with periodic boundary conditions, and use the system of units in which $\hbar\equiv1$ and $m\equiv1/2$. Note that we have introduced the parameters $N$ and $R$ merely to define the model. In what follows, these parameters do not play a role: We calculate many-body properties either analytically in the thermodynamic limit ($N,R\to\infty$ with the fixed Fermi energy) or numerically with (sufficiently) large values of $N$ and~$R$.

The ground state of Hamiltonian~\eqref{eq_hamiltonian_heavy} is the Slater determinant $\Psi(\bm{r}_1,...,\bm{r}_N)=(N!)^{-1/2}\det[\psi_j(\bm{r}_i)]_{i,j\leq N}$ over single-body wave functions $\psi_j(\bm{r}_i)$ that satisfy the Schr{\"o}dinger equation
\begin{equation}
-\Delta_i\psi_j(\bm{r}_i)+U(\bm{r}_i)\psi_j(\bm{r}_i) =\varepsilon_j \psi_j(\bm{r}_i)\,,
\label{eq_schroedinger}
\end{equation}
where $\varepsilon_j$ denotes the energy of the $j$th fermionic state. The states are ordered such that $\varepsilon_1\leq \varepsilon_2\leq\cdots\leq\varepsilon_N$. Note that the subscript $i$ does not carry any important information, so we shall omit it whenever possible.  

For later convenience, we introduce the Fermi energy of the bath as
\begin{equation}
\varepsilon_F\equiv\varepsilon_N\,.
\end{equation}
The Fermi momentum, $k_F$, is defined via $k_F^2\equiv \varepsilon_F$.
$k_F$ and $\varepsilon_F$ set length and energy scales in our problem. To illustrate our findings, we shall plot dimensionless quantities that correspond to the system of units in which $k_F=1$.     

Further analysis of the problem depends on the definition of the potential $U$, which incorporates information about the internal structure of the impurity. We focus on the $s$-wave interaction and assume that there is no interaction in the higher partial waves: this is a standard approximation for cold-atom systems.  In other words, we assume that the impurity-fermion interaction is radial $U(\bm{r})\equiv U(r)$ and acts only on the wave functions with zero angular momentum. This assumption allows us to work with the $s$-wave radial Schr{\"o}dinger equation 
\begin{equation}
-\frac{\mathrm{d}^2}{\mathrm{d}r^2}\phi_{j} (r)+U(r)\phi_{j} (r)=k_j^2 \phi_{j} (r)\,,
\label{eq_radial}
\end{equation}
where $\phi_{j}(r)\equiv r \psi_{s,j}(r)$  [$\psi_{s,j}(r)$ is $\psi_j(\bm{r})$ for $s$ waves].  The momentum of a fermion, $k_j$, is defined via the relation $k_j^2\equiv\varepsilon_j$. 

Equation~\eqref{eq_radial} is a textbook one-body problem. Since we are interested in the physics outside the range of the potential $r_I$, it is natural to define $U(r)$ through its scattering properties, namely, the \textit{phase shift} $\delta(k)$, which defines the solution $\phi_j(r)$ at $r>r_I$~\cite{Scattering_Taylor}:
\begin{equation}
\phi_{j}(r>r_I) = \gamma_j \sin[k_j r+\delta(k_j)]\,,
\label{eq_phase_shift_wave_function}
\end{equation}
where $\gamma_j$ is the normalization coefficient. The function in
Eq.~\eqref{eq_phase_shift_wave_function} is a linear combination of the solutions to Eq.~\eqref{eq_radial} with $U(r)=0$. Without the impurity only the regular solution is possible,~i.e.,~$\delta(k)\equiv0$. The presence of the impurity induces a non-zero phase shift that contains information about the impurity. 

Note that the introduced model cannot be used to model the behavior of the Fermi gas inside the impurity potential, i.e., if $r<r_I$. However, one must take into account that the effective range of atom-atom potentials, $r_I$, is typically on the nanometer scale, whereas the length scale associated with $1/k_F$ is of the order of a micron. This separation of scales implies that our model correctly captures the physics relevant for cold-atom experiments assuming that an impurity is an atom.

%%%%%%%%%%%%%%%%%%%%%%%%%%%%%%%%%%%%%%%%%%%%%%%%%%
\begin{figure}
	\centering
	\includegraphics[width=\linewidth]{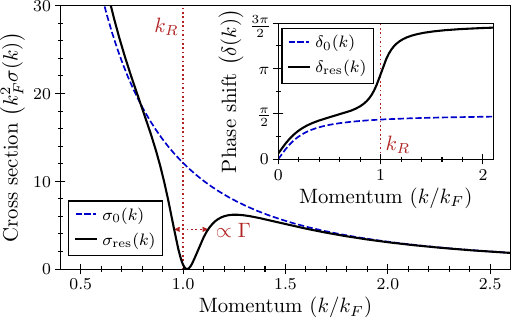}
	\caption{(main panel) The solid black curve illustrates the dependence of the scattering cross section $\sigma_{\text{res}}(k)$ on the momentum $k$~[see Eq.~\eqref{eq_optical_theorem}]. The cross section exhibits a resonance at $k\approx k_{R}$ (dotted vertical line).
	The width of the resonance is determined by $\Gamma$. For comparison, we also present $\sigma_0(k)=\frac{4\pi}{k^{2}}\sin^2\big(\delta_0(k)\big)$ (dashed blue curve).
		(inset)~The scattering phase shift as a function of the momentum. Note that $\delta_{\text{res}}(k)$ changes rapidly in the vicinity of the resonance. Far from the resonance, both the phase shift and the cross section are determined by $\delta_0(k)$. We assume strong background interaction with $k_F\alpha=-5$. The parameters of the resonance are $\Gamma=0.2\,\varepsilon_{F}$ and $k_R=k_F$.}
	\label{fig_system}
\end{figure}
%%%%%%%%%%%%%%%%%%%%%%%%%%%%%%%%%%%%%%%%%%%%%%%%%%%%%%%%%%%%%%%%%%%%%%%%%%

 We use a simple model of an impurity with a \textit{resonant} scattering phase shift. Namely, we consider $\delta_{\text{res}}(k)$ of the following form~\cite{Scattering_Taylor}:
\begin{equation}
\delta_{\text{res}}(k)=\delta_\text{0}(k)+\atan(\frac{k^{2}-k_{R}^2}{\Gamma})+\frac{\pi}{2}\,,
\label{eq_phase_shift}
\end{equation}
where $\Gamma$ is the characteristic width of the resonance and $k_{R}$ is the resonant momentum. Such a phase shift exhibits an abrupt change by $\pi$ in the vicinity of the critical point $k\approx k_R$ (see the inset of Fig.~\ref{fig_system}). 

To simplify analytical calculations in the following sections, we also introduce the approximation $\delta^{*}_{\text{res}}(k)$ to the phase shift~\eqref{eq_phase_shift} that is accurate for narrow resonances $(\Gamma\to 0)$:
\begin{equation}
	\delta_{\text{res}}(k)\approx\delta^{*}_{\text{res}}(k)= \delta_\text{0}(k)+\atan(\frac{k-k_{R}}{\omega/2})+\frac{\pi}{2}\,,
	\label{eq_phase_shift_approximate}
\end{equation}
where $\omega\equiv \Gamma/k_R$.

The precise form of $\delta_0(k)$ is not important for our analysis, as long as it does not change rapidly as a function of $k$. For the sake of discussion, we use 
\begin{equation}
\delta_\text{0}(k)\equiv-\atan(k\alpha),
\label{eq_phase_shift_simple}
\end{equation} 
where $\alpha$ is the scattering length. This form of phase shift is typical for low-energy scattering, in particular, it is the standard form for scattering of cold atoms~\cite{Feshbach_review}. However, even for cold atoms, when the width of a characteristic Feshbach resonance is much smaller than the Fermi energy, one needs to consider more complicated forms of the phase shift, e.g., the one in Eq.~(\ref{eq_phase_shift}), see Ref.~\cite{Feshbach_review}. 
The scattering length $\alpha$ can be either negative or positive. In the latter case, the potential $U$ can support shallow bound states. We assume $\alpha<0$ to focus only on the physics associated with the resonance. Nevertheless, it is straightforward to extend our approach and results to the case with positive scattering lengths.

The scattering cross section, $\sigma(k)$, for the fermion-impurity collision reads as
\begin{equation}
\sigma_{\mathrm{res}}(k)=\frac{4\pi}{k^{2}}\sin^2\big(\delta_{\mathrm{res}}(k)\big)\,.
\label{eq_optical_theorem}
\end{equation}
It exhibits a resonance with a Breit-Wigner profile~\cite{Breit_Wigner} (see Fig.~\ref{fig_system}). The value of the scattering length $\alpha$ defines whether $\sigma_{\mathrm{res}}$ exhibits a peak $(|k_F\alpha|\gg 1)$ or a dip $(|k_F\alpha|\ll1)$~\cite{Scattering_Taylor}. To illustrate our findings graphically, we shall use a large impurity-fermion scattering length, $k_F\alpha=-5$. This allows us to provide an insight into the near-to-the-unitarity regime, which is usually challenging from the theoretical point of view. 

The main advantage of the simple interaction model presented above is that it allows us to access analytically the many-body properties of the system. Additionally, the aforementioned model can be easily extended to impurities with a discrete internal spectrum with more than a single level. This said, present studies might be beneficial for the prospective research on an analog of the angulon quasiparticle~\cite{Angulon_1,Angulon_2} in a fermionic environment.

\section{Self-Energy of the Impurity}
\label{sec_energy}

The typical starting point for an analysis of systems with impurities is a calculation of the energy spectrum. A quantity of particular interest is the amount of energy that is required to immerse the impurity in the Fermi gas. It is defined as the difference between the ground-state energies of the Fermi gas with and without the impurity. In the language of Dyson's equation, this energy is determined by the sum of the self-energies of the fermions in the presence of the impurity. By analogy, we shall call this quantity the self-energy of the impurity, $\varepsilon_{I}$. Note that $\varepsilon_{I}$ is of direct experimental importance, as it determines the onset of the excitation branch in radio-frequency spectroscopy, which is one of the standard tools for studying impurities in cold Fermi gases~(for review see Ref.~\cite{Heavy}).

To calculate $\varepsilon_{I}$, we employ  
Fumi's theorem~\cite{Mahan,Fumi}, which holds exactly in our model. The theorem states that $\varepsilon_{I}$ in the thermodynamic limit is a simple integral that involves only the scattering phase shift $\delta(k)$:
\begin{equation}
	\varepsilon_{I}=-\frac{2}{\pi}\int\limits_{0}^{k_F}k\delta(k)\mathrm{d}k\,.
	\label{eq_Fumi}
\end{equation}
Note that the scattering phase shift is positive~(see Fig.~\ref{fig_system}), which implies $\varepsilon_{I}<0$.

For the non-resonant phase shift $\delta_0(k)$, the integral in Eq.~\eqref{eq_Fumi} leads to a simple expression (cf.~\cite{Chevy_Fermi_polaron})
\begin{equation}
	\varepsilon_{I,0}=-\frac{\varepsilon_{F}}{\pi}\bigg[y+(1+y^2)\Big(\frac{\pi}{2}+\atan(y)\Big)\bigg]\,,
	\label{eq_energy_impurity}
\end{equation}
where $y\equiv1/(k_F\alpha)$. For a system with a resonance, this expression defines the base line of $\varepsilon_{I}$.

\begin{figure}
	\centering
	\includegraphics[width=\linewidth]{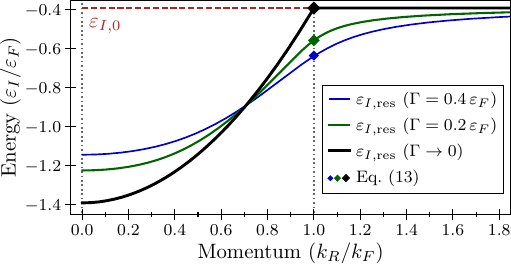}
	\caption{The energy $\varepsilon_{I,\text{res}}$ as a function of the position of the resonance, $k_R$ (solid curves). For $k_R\gg k_F$, the energy of the system with a resonance converges to the value in Eq.~\eqref{eq_energy_impurity}, which is shown as a dashed red line. For $k_R\leq k_F$, the energy is decreased compared to $\varepsilon_{I,0}$, due to an extra energy level available for fermions. At $k_R=k_F$ the energy can be accurately approximated by Eq.~\eqref{eq_energy_approximate} whose predictions are demonstrated by markers. We use $k_F\alpha=-5$.}
	\label{fig_energy}
\end{figure}

In general, the integral in Eq.~\eqref{eq_Fumi} leads to a cumbersome expression, which we do not present here. Instead, we calculate the self-energy numerically. In Fig.~\ref{fig_energy}, we plot the dependence of $\varepsilon_{I,\text{res}}$ on the momentum $k_{R}$ for different values of the width parameter $\Gamma$. A resonance far outside the Fermi sphere $(k_R\gg k_F)$ cannot affect the system, and the energy converges to the value determined by Eq.~\eqref{eq_energy_impurity}.

If the resonance is inside the Fermi sphere $(k_R<k_F)$, a fermionic state quasibound to the impurity becomes accessible. The energy of this state is lower than $\varepsilon_F$, and therefore, after the rearrangement of fermions, $\varepsilon_{I,\text{res}}$ becomes smaller than $\varepsilon_{I,0}$. This can be most easily understood in the limit $\Gamma\to0$. In this limit the resonance can be treated as a single level with the energy $k^2_R$. The rearrangement of the Fermi gas means here that one fermion that was at the Fermi surface is pushed to the resonance level. The associated energy shift, and hence $\varepsilon_{I,\text{res}}-\varepsilon_{I,0}$, is given by $k^2_R-k^2_F<0$ (see Fig.~\ref{fig_energy}, solid black curve).

All curves in Fig.~\ref{fig_energy} intersect at one point independent of other parameters. This occurs when $k_R^2=k_F^2/2$, i.e., when the resonance energy is exactly the half of the Fermi energy. In this case, the part of the integral in Eq.~\eqref{eq_Fumi} that contains $\atan[(k^2-k^2_R)/\Gamma]$ vanishes, leading to $\varepsilon_{I,\text{res}}(k_R^2=k_F^2/2)=\varepsilon_{I,0}-k_F^2/2$.

%%%%%%%%%%%%%%%%%%%%%%%%%%%%%%%%%%%%%%%%%%%%%%%%%%
%%%%%%%%%%%%%%%%%%%%%%%%%%%%%%%%%%%%%%%%%%%%%%%%%%%%

As we focus on the regime with $k_R\approx k_F$, and $\Gamma\to0$, 
we can use the approximation to the phase shift $\delta_{\text{res}}^{*}(k)$ to estimate the self-energy:
\begin{align}
&\varepsilon_{I,\text{res}}\approx\varepsilon_{I,0}-\frac{\varepsilon_F}{2}+\frac{k_F \omega}{2\pi}-\frac{\varepsilon_F}{\pi}\atan(\frac{k_F-k_R}{\omega/2})\nonumber\\
&+\frac{i}{2\pi}\Bigg[\chi_{-}^2\ln(1+\frac{k_F}{\chi_{-}})-\chi_{+}^2\ln(1-\frac{k_F}{\chi_{+}})\Bigg]\,,
\label{eq_energy_resonance}
\end{align}
where $\chi_{\pm}\equiv i\omega/2\pm k_{R}$. Similar to Eq.~\eqref{eq_phase_shift_approximate}, this approximation is accurate, when $\omega\to 0$. If the resonance is directly at the Fermi surface $(k_R=k_F)$, the energy $\varepsilon_{I,\text{res}}$ is approximated by (see the markers in Fig.~\ref{fig_energy})
\begin{equation}
\varepsilon_{I,\text{res}}\big|_{k_R\to k_F}\approx\varepsilon_{I,0}-\frac{\Gamma}{\pi}\ln(\frac{\Gamma}{e \varepsilon_{F}})\,.
\label{eq_energy_approximate}
\end{equation}
Note that this expression is non-analytical in $\Gamma$, which indicates that one has to be careful when using perturbative approaches to study the problem. The non-analytical behavior of the energy also suggests the emergence of a relevant length scale in addition to the one given by the density of the Fermi gas. That length scale should be proportional to $1/\Gamma$ as is evident from the argument of the logarithm in Eq.~(\ref{eq_energy_approximate}). For $\Gamma\to 0$, it should determine the physics of the system far outside the size of the potential. In order to verify this, the next step of our investigation is to consider the density profile of the Fermi gas.

\section{Density of the Fermi gas}
\label{sec_density}

The density of the Fermi gas can be written as
\begin{align}
	n(r) = \rho_{l=0}(r)+\rho_{l>0}(r)\,,
\end{align}
where $\rho_{l=0}(r)\equiv\rho(r)/r^2$ with $\rho(r)\equiv\sum_{i=1}^{N_s} \big|\phi_{i}(r)\big|^2$, where $N_s$ is the number of $s$-wave fermions. The $\rho_{l>0}(r)$ describes the density due to the higher angular momenta. It is not affected by the impurity, hence, we do not discuss it from now on. In the thermodynamic limit, the summation over single-particle fermionic states can be replaced by the integration over momenta so that
\begin{equation}
	\rho(r) = \frac{k_F}{\pi}-\frac{1}{\pi}\int\limits_0^{k_F}\cos\big(2kr+2\delta(k)\big)\mathrm{d}k\,.
	\label{eq_density}
\end{equation}
In the limit $r\to\infty$, the function $\cos\big(2kr+2\delta(k)\big)$ oscillates so rapidly that it contributes to the integration only over an incomplete period directly at the Fermi surface. Therefore, for any scattering phase shift $\delta(k)$, we can write 
\begin{equation}
	\Delta \rho(r\to\infty)\to\Delta \rho_{\text{uni}}(k_F,r)\equiv-\frac{\sin\big(2k_Fr+2\delta(k_F)\big)}{2r}\,,
	\label{eq_universal}
\end{equation}
where we define $\Delta \rho(r)\equiv \pi\rho(r)-k_F$. This long-range behavior of the density function is called the Friedel oscillations~\cite{Friedel}. Its universality is not only of a theoretical interest; it also suggests observables for understanding the many-body environment~\cite{Friedel_bath_1,Friedel_bath_2,Friedel_bath_3} or the defect itself~\cite{Friedel_impurity}. 
 In general, any noticeable modification in the behavior of the density oscillations can be used as a probe of the system. 

For impurities without any internal structure, Eq.~(\ref{eq_universal}) describes the density well even in the vicinity of the impurity. However, the resonance can strongly affect the density, and additional terms should be added to Eq.~(\ref{eq_universal}) for a faithful description of the density profile. To understand this, note that one can use $\delta(k_F)$ in the argument of Eq.~(\ref{eq_density}) only if $\delta(k_F)\simeq \delta(k_F\pm1/r)$. With a resonance at the Fermi level, this condition means that the Friedel oscillations do not describe the density profile if $r\lesssim k_F/\Gamma$.

%%%%%%%%%%%%%%%%%%%%%%%%%%%%%%%%%%
\begin{figure}
	\centering
	\includegraphics[width=\linewidth]{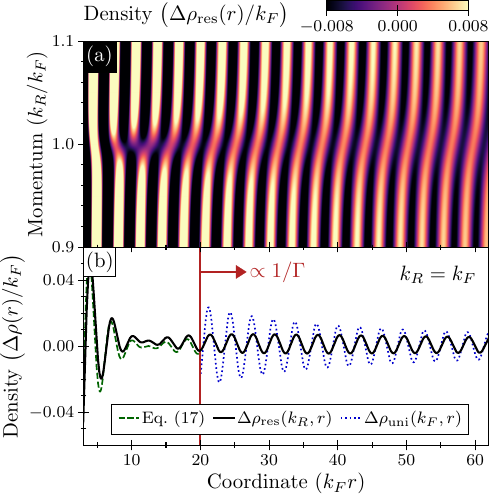}
	\caption{(a)~Dependence of the density, $\Delta \rho_{\text{res}}(r)$, on the resonance momentum $k_R$ and the coordinate $r$. Near the critical point $k_R\approx k_F$, the density is strongly modified~(see the text for details). (b)~Density in the presence of the impurity with the resonant scattering phase shift $\delta_{\text{res}}(k)$~(solid black curve). The resonance is at the Fermi surface $(k_R=k_F)$. Far from the impurity, density oscillations converge to the universal behavior~(dotted blue curve). The oscillations close to the impurity ($r\lesssim k_F/\Gamma$) can be approximated by Eq.~\eqref{eq_density_approximation}~(dashed green curve).   We use $k_F\alpha=-5$, $r_I=0$, and $\Gamma=0.05\,\varepsilon_{F}$.}
	\label{fig_density}
\end{figure}
%%%%%%%%%%%%%%%%%%%%%%%%%%%%%%%%

To illustrate this argument,
 we compute $\Delta \rho_{\mathrm{res}}(r)$~(see Fig.~\ref{fig_density}). Figure~\ref{fig_density}(a) shows the density for different values of $k_R$. We see that if $k_R$ is far from $k_F$, i.e., $|k_R^2-k_F^2|>\Gamma$, the Friedel oscillations describe the density well for almost any value of $r$. However, if $|k_R^2-k_F^2|<\Gamma$, then the resonance  strongly modifies the density profile~[see also Fig.~\ref{fig_density}(b) where $k_R=k_F$]. Our conclusion here is that far from the impurity, the density always exhibits the Friedel oscillations. However, near the defect, the density can be non-trivially deformed. 
 Note that $k_R/\Gamma$ is approximately $4\,\mu$m for a bath of $^6$Li atoms with $T_F\sim1\,\mu$K and $\Gamma=0.1\,\varepsilon_F$. Observation of density modulations on such length scales should be possible via quantum gas microscopy~\cite{Riechers2017}.

The behavior of the density presented in Fig.~\ref{fig_density} can be connected to node counting in Friedel's sum rule (for an introduction to the rule, see, e.g.,~\cite{Coleman}). If $k_R\gg k_F$, the number of full oscillations of $\rho(r)$ within the range $r\in[r_I,R]$ should be equal to the number of $s$-wave fermions $N_s$ in the bath. If the resonance
lies inside the Fermi sea, the fermions ``rearrange'' and the number of nodes in $r\in[r_I,R]$ is $N_s-1$. To illustrate this, we calculate the density, $\Delta \rho_{\text{res}}(r)$, for different values of $k_R$~[see Fig.~\ref{fig_density}(a)]. The profile of the density reveals the deformation for $|k_F-k_R|<\sqrt{\Gamma}$, i.e., when the fermions at the Fermi level ``rearrange''.

To provide further analytical insight into the functional form of the density when $k_R\approx k_F$, we use the approximate form of the phase shift~\eqref{eq_phase_shift_approximate} with $\delta_{0}(k)\to\delta_{F}\equiv \delta_0(k_F)$. In what follows we also assume that $\delta_0(k_R)=\delta_0(k_F)$. Such a simplification allows us to approximate the density as follows:
\begin{equation}
\Delta \rho_{\text{res}}(r)\approx \Delta  \rho_{\text{def}}(r)-\Delta \rho_{\text{uni}}(k_F,r)\,,
\label{eq_density_approximation}
\end{equation}
where 
\begin{align}
\Delta  \rho_{\text{def}}&(r)=\omega e^{\omega r}\Big[\cos\big( 2k_{R}r+2\delta_F\big)\Im\big(\mathcal{J}_{+}(r)\big)\nonumber\\&+\sin\big( 2k_{R}r+2\delta_F\big)\Re\big(\mathcal{J}_{-}(r)\big)\Big] \nonumber\\&\approx \omega e^{\omega r}\sin\big( 2k_{R}r+2\delta_F\big)\mathrm{Ei}(-\omega r)\,,
\end{align}
and
\begin{equation}
\mathcal{J}_{\pm}(r)\equiv \mp\mathrm{E}_{1}\big[-2i\chi_{+}r\big]-\mathrm{E}_{1}\big[-2i(\chi_{+}-k_{F})r\big]\,.
\end{equation}
$\mathrm{E}_1(z)$ is the generalization of  the exponential integral $\mathrm{Ei}(x)$ on the complex plane~\cite{Abramowitz_Stegun}. Approximation~\eqref{eq_density_approximation} is accurate  when $\omega\to 0$,~i.e., for narrow resonances, see Fig.~\ref{fig_density}(b). It reveals that the deformation of the many-body density function is induced by an effective interplay of the oscillations with periods given by $k_R$ and $k_F$.

We use the asymptotic expansion of the exponential integral $\mathrm{Ei}(-\omega r)$ and derive the density for $r\to\infty$:
\begin{equation}
	\Delta \rho_{\text{res}}(r)\approx-\frac{\sin\big(2k_F r+2\delta_F \big)}{2r}+\frac{\sin\big(2k_R r+2\delta_F\big)}{\omega r^2}\,.
	\label{eq:delta_rho_expansion}
\end{equation}
This expression shows that the size of the deformed region is inversely proportional to the width parameter of the resonance $\Gamma$.  For $r\gg k_R/\Gamma$, the second term becomes irrelevant and $\Delta \rho_{\text{res}}(r)$ converges to the universal Friedel oscillations~\eqref{eq_universal}. Note that the expression in Eq.~(\ref{eq:delta_rho_expansion}) is universal, in a sense that it depends only on the parameters of the resonance, and not on the short-range physics.

To summarize this section: A narrow resonance in the vicinity of the Fermi energy  introduces a length scale given by $k_R/\Gamma$, which leads to the density with two distinct patterns. As $r\to\infty$, the Friedel oscillations describe the density well. However, at distances $r\simeq k_R/\Gamma$, there is an additional oscillatory term. The behavior that we observed has some similarities with those of the Kondo model where the extra length scale is given by the Kondo temperature~(see also a similar observation in Ref.~\cite{Tao2012}). In particular, the charge density in the Kondo model features a crossover from the short- to long-distance regimes, which cannot be simply approximated by the Friedel oscillations~\cite{Affleck2008} [cf.~the interplay between the terms proportional to $1/r$ and $1/r^2$ in Eq.~(\ref{eq:delta_rho_expansion})].

\section{Dynamics upon immersion of the impurity}
\label{sec_QS}

So far we have studied only the ground-state properties.  In this section, we focus on the corresponding time-dependent problem, which helps us to understand how the system reaches equilibrium following an immersion of the impurity. In addition, quench dynamics allows us to visualize relevant time scales in impurity models, which can be observed in experiments sensitive to real-time evolution~(see, e.g., Ref.~\cite{Fermi_polaron_experiment_3}).

%%%%%%%%%%%%%%%%%%%%%%%%%%%%%%%%%%
\begin{figure}
	\centering
	\includegraphics[width=\linewidth]{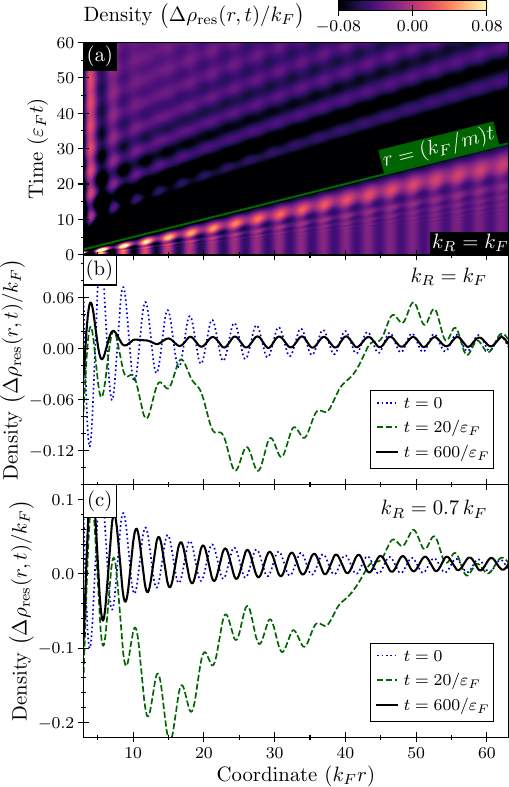}
	\caption{(a)~Time evolution of $\Delta \rho_{\text{res}}(r,t)$ with $k_R=k_F$ after a sudden immersion of the impurity at $t=0$. The deformation of the density propagates with the group velocity $k_F/m$. (b),~(c)~At large times $(t\to\infty)$, the density converges to the result discussed in Sec.~\ref{sec_density}. If $k_R=k_F$, the density oscillations are deformed~(solid curve) when compared to the universal Friedel oscillations~(dotted curve). The standard Friedel oscillations are realized if the resonance is far from the Fermi level, here with $k_R=0.7\,k_F$. In numerical simulations, we use $k_F\alpha=-5$, $r_I=0$ and $\Gamma=0.05\,\varepsilon_{F}$.}
	\label{fig_time_dynamics}
\end{figure}
%%%%%%%%%%%%%%%%%%%%%%%%%%%%%%%%%%%%%

We consider the following quench protocol: at $t<0$ ($t$ for time) there is no fermion-impurity interaction. At $t=0$, the impurity-fermion interaction, $U(r)$, is turned on, which corresponds to the immersion of the defect into the bath. We analyze time evolution of the composite system at $t>0$. As before, we focus only on the behavior of the $s$ waves.

At $t=0$, the wave function of the Fermi gas is the Slater determinant over the non-interacting functions $\phi^{(0)}_{j}(r)=\sqrt{2/R}\sin (\kappa_j r)$, 
where the momenta $\kappa_j\equiv\pi j/R$ are determined from the boundary condition $\phi^{(0)}_{j}(R)=0$. In other words, we have 
\begin{equation}
	\Phi(r_1,...,r_{N_s}; t=0)=\frac{1}{\sqrt{N_s!}}\det\big[\phi^{(0)}_{j}(r_i)\big]_{i,j\leq N_s}\,.
	\label{eq:Psi_0_det}
\end{equation}
The corresponding density is
\begin{equation}
	\rho(r,t=0)=\frac{2}{R}\sum\limits_{i=1}^{N_s} \sin[2](\kappa_i r)\,.
\end{equation}

At $t>0$, time evolution of the wave function $\Phi(r_1,...,r_{N_s},t)$ can be formally written as
\begin{equation}
	\Phi(r_1,...,r_{N_s};t)=\exp(-i\hat{\mathcal{H}}_{s}t)\Phi_0(r_1,...,r_{N_s}; t=0)\,,
	\label{eq_time_evolution}
\end{equation}
where $\hat{\mathcal{H}}_{s}$ is the Hamiltonian of the $s$-wave part of the system. In order to simplify the evaluation of Eq.~\eqref{eq_time_evolution}, we expand the single-particle wave functions $\phi^{(0)}_{j}$ that enter Eq.~(\ref{eq:Psi_0_det}) in the basis of one-body wave functions $\phi_{j}$ (see Eq.~(\ref{eq_radial}))
\begin{equation}
	\phi^{(0)}_{j}(r_i)=\sum\limits_{j'=1}^\infty A_{jj'}\phi_{j'}(r_i)\,,
\end{equation}
where the expansion coefficients are defined as
\begin{equation}
	A_{jj'}=\int\limits_0^R\phi^{(0)}_{j}(x)\phi_{j'}(x)\,\mathrm{d}x\,.
\end{equation}

Then, the density at $t>0$ reads as follows
\begin{align}
	\rho (r,t)=&\int\Phi^{*}(r,r_2,...,r_{N_s},t)\Phi(r,r_2,...,r_{N_s},t)\,\mathrm{d}r_2...\mathrm{d}r_{N_s}\nonumber\\
	&=\sum\limits_{j=1}^{N_s} \Bigg|\sum\limits_{j'=1}^\infty \exp(-i\varepsilon_{j'}t)A_{jj'}\phi_{j'}(r)\Bigg|^2\,.
	\label{eq_density_dynamic}
\end{align}
We illustrate this density in Fig.~\ref{fig_time_dynamics} for a bath with $N_s=300$, $R=\pi N_s/k_F$ and $k_R=k_F$. The momentum cutoff, which is required for the numerical evaluation of Eq.~\eqref{eq_density_dynamic}, is set to $k_\infty=20\,k_F$. At $t>0$, the density begins to deform. The deformation occurs in the light cone determined by the group velocity $k_F/m$ (see Fig.~\ref{fig_time_dynamics}(a)) --  similar to the result of Ref.~\cite{Friedel_dynamic} without a resonance. Such a behavior is typical also for other models with impurities~\cite{Medvedyeva2013,Lechtenberg2014}.

Our numerical simulations do not show clear signs of a new time scale in the time dynamics of the density of the Fermi gas. The presence of the resonance is apparent only in the long-time limit $(t\to\infty)$, as exemplified in Figs.~\ref{fig_time_dynamics}(b)~and~\ref{fig_time_dynamics}(c) for $t=600/\varepsilon_F$, which corresponds to experimentally feasible 29 ms in a bath of $^6$Li atoms at $T_F\sim1\,\mu$K. In particular, the oscillations of the density for $k_R=k_F$ are deformed in comparison to the Friedel oscillations that occur when the resonance is located far from the Fermi energy,~e.g., at~$k_R=0.7\,k_F$. Note that in cold-fermion systems, characteristic time scales of the order of hundred $1/\varepsilon_F$ allow a system to reach equilibrium upon immersion of a heavy impurity~(see, e.g.,~Ref.~\cite{Heavy}). In our study, we also observe that the results at these time scales agree with the ground-state calculations presented in the previous section.

\section{Finite-temperature properties}
\label{sec_temp}

Density oscillations due to the presence of the resonance can be observed experimentally. For example, in cold-atom systems, they could be detected either using another static impurity as a probe~\cite{Recati2005} or a quantum gas microscope~\cite{Riechers2017}. To investigate the feasibility of such a measurement, we study the system at finite temperatures. Note that typical temperatures in current cold-atom experiments are $\tau\simeq 0.1$, where $\tau\equiv k_{B}T/\varepsilon_{F}$ and $k_{B}$ is the Boltzmann constant~(see Ref.~\cite{Giorgini2008} and references therein).

%%%%%%%%%%%%%%%%%%%%%%%%%%%%%%%%%%
\begin{figure}
	\centering
	\includegraphics[width=\linewidth]{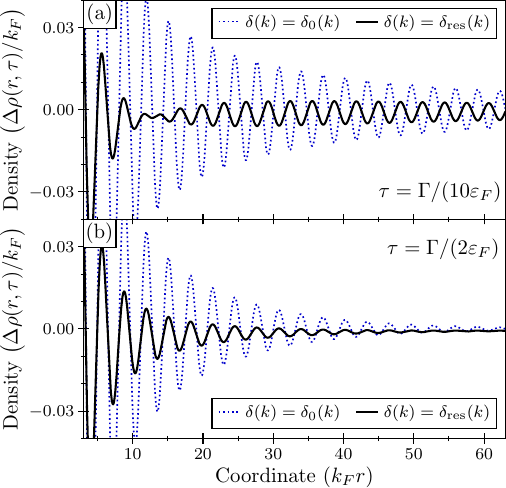}
	\caption{The density $\Delta \rho(r,\tau)$ of the Fermi gas in the presence of the impurity with the resonant phase shift $\delta_{\text{res}}(k)$~(solid black curve) or with the phase shift $\delta_{0}(k)$~(dotted blue curve). We consider the resonance at the Fermi surface $(k_R=k_F)$. (a)~At temperatures much lower than $\Gamma/k_B$, the density is approximately given by the ground-state wave function. (b)~At higher temperatures, the density is strongly affected by the temperature. We use $k_F\alpha=-5$ and $\Gamma=0.05\,\varepsilon_{F}$.}
	\label{fig_finite_temperature}
\end{figure}
%%%%%%%%%%%%%%%%%%%%%%%%%%%%%%%%

 The system in thermodynamic equilibrium at finite temperatures is described by the Fermi-Dirac distribution  
\begin{equation}
n_{\text{FD}}(k,\tau)=\frac{1}{1+\exp(\frac{k^2}{k_{F}^2\tau}-\tilde{\mu}(\tau))}\,,
\end{equation}
where the chemical potential, $\tilde{\mu}(\tau)\equiv\mu/\varepsilon_{F}$, is given by~\cite{Cowan2019}
\begin{equation}
\tilde{\mu}(\tau)=\tau \log(-\mathrm{Li}_{3/2}^{-1}\bigg(-\frac{4}{3\sqrt{\pi}}\tau^{-3/2}\bigg))\,.
\end{equation}
Here, $\mathrm{Li}_{3/2}^{-1}(x)$ is the inverse of the polylogarithm function. At low temperatures $(\tau\ll 1)$, one can can use an approximate expression $\mu\simeq 1-\pi^2\tau^2/12$.

 To study the effect of the temperature on the density, we consider the function 
\begin{equation}
\Delta\rho(r,\tau) = \int\limits_0^{\infty}\cos\big(2kr+2\delta(k)\big)n_{\text{FD}}(k,\tau)\mathrm{d}k\,,
\label{eq_density_thermal}
\end{equation}
which contains all information about the density oscillations.
The dependence of $\Delta\rho(r,\tau)$ on the coordinate $r$ is illustrated in Fig.~\ref{fig_finite_temperature}.  We compare the density profiles without (dotted curves) and with (solid curves) a resonance. In the latter case, we assume that $k_R=k_F$. At temperatures significantly lower than $\Gamma/k_B$,~e.g., at~$\tau=\Gamma/(10\,\varepsilon_{F})$, the ground-state result is accurate~[see Figs.~\ref{fig_finite_temperature}~(a)~and~\ref{fig_density}]. In contrast, at the temperatures comparable to the width of the resonance,~e.g., at~$\tau=\Gamma/(2\,\varepsilon_{F})$, the effect of the temperature is noticeable~[see Fig.~\ref{fig_finite_temperature} (b)], especially for a system with a resonance. 
In general, the difference between systems with and without a resonance becomes less prominent when increasing the temperature; it will disappear at temperatures $\tau\gg \Gamma/\varepsilon_F$. Indeed, the density oscillations are induced by the sharp edge of the Fermi sphere; they disappear at higher temperatures when the energy distribution of fermions becomes more gradual.       
We conclude that the resonance introduces a temperature scale into the problem, and only if $\tau\lesssim \Gamma/\varepsilon_{F}$ one can clearly see the effect of the resonance.  Assuming a narrow resonance, $\Gamma/\varepsilon_{F}\simeq 0.1$, this condition implies that our findings are within reach of the state-of-the-art cold-atom experiments.

It is worth noting that one can also calculate the self-energy of the impurity $\varepsilon_{I}(\tau)$ at finite temperatures. To this end, one should use the extended version of Fumi's theorem (cf.~\cite{Liu2020})
\begin{equation}
	\varepsilon_{I}(\tau)=-\frac{2}{\pi}\int\limits_{0}^{\infty}k\delta(k)n_{\text{FD}}(k,\tau)\mathrm{d}k\,.
	\label{eq_Fumi_finite_temperature}
\end{equation} 
For the temperatures of our interest $(\tau\lesssim 0.1)$, Eq.~\eqref{eq_Fumi_finite_temperature} does not produce values significantly different from Eq.~\eqref{eq_Fumi}. This is expected: the energy is not sensitive to weak thermal perturbations at the surface of the Fermi sphere. Therefore, we refrain from discussing the self-energy of the impurity at finite temperatures further.

\section{Implications for the Fermi polaron}
\label{sec_MF}

Finally, we use our findings in the context of another important model in cold-atom physics -- the Fermi polaron~\cite{Chevy2006,Chevy_Fermi_polaron,Svistunov2008}, which is a quasiparticle introduced to describe experiments with two-component Fermi gases that have large (quasi)-spin imbalance~\cite{Fermi_polaron_experiment_1,Fermi_polaron_experiment_2,Fermi_polaron_experiment_3,Scazza2017} (for review, see Refs.~\cite{Fermi_polaron_review,Fermi_polaron_theory_review}). 
First of all, it is clear that our results cannot be easily extended to describe the residue or long-range spatial profile (e.g., the impurity-fermion correlation function) of the Fermi polaron. These properties are non-analytical in the mass of the impurity. In particular, they acquire important logarithmic corrections for large masses~[see, e.g.,~Ref.~\cite{Trefzger2013}]. 
By contrast, the self-energy is not expected to drastically change for the impurity with a finite mass~\cite{Chevy_Fermi_polaron}. Here, we estimate this energy using our results for a static impurity.
For a different perspective on Fermi polarons near narrow Feshbach resonances, see Refs.~\cite{Trefzger2012,Massignan2012,Qi2012}.

The Hamiltonian of a mobile impurity in a Fermi gas is similar to Eq.~\eqref{eq_hamiltonian_heavy} and reads as
\begin{equation}
\hat{\mathcal{H}}_{\text{pol}}=-\frac{\hbar^2}{2M}\Delta_{0}-\frac{\hbar^2}{2m}\sum\limits_{i=1}^{N}\Delta_i+\sum\limits_{i=1}^{N}U(\bm{r}_i-\bm{r}_0)\,,
\label{eq_hamiltonian_polaron}
\end{equation}
where the subscript $0$ refers to the impurity with mass $M$. As in Sec.~\ref{sec_resonance}, we assume periodic boundary conditions at $|\bm{r}|=R$, and work in a system of units with $m\equiv1/2$ and $\hbar\equiv1$.  

To study the Hamiltonian~\eqref{eq_hamiltonian_polaron}, we first note that its eigenstates can be written as 
\begin{equation}
\Psi(\bm{r}_0,...,\bm{r}_N)=\sum_{\bm{k}_0,...,\bm{k}_N}A(\bm{k}_0,...,\bm{k}_N)\prod\limits_{j=0}^N e^{-i \bm{k}_j \bm{r}_j},
\end{equation}
where the wave vectors are quantized as $\bm{k}=\pi\bm{n}/R$ with integers $\bm{n}\equiv\{n_x,n_y,n_z\}$. The potential $\sum_{i=1}^N U(\bm{z}_i)$ depends only on the relative coordinates $\bm{z}_i\equiv \bm{r}_i-\bm{r}_0$, which means that the total momentum of the system ($\bm{Q}\equiv\sum_{i=0}^N \bm{k}_i$) is conserved, and that the wave function can be written as 
\begin{equation}
\Psi(\bm{r}_0,...,\bm{r}_N)=e^{-i\bm{Q}\bm{r}_0}\Phi(\bm{z}_1,...,\bm{z}_N)\,.
\label{eq:LLP_phi}
\end{equation}
This observation motivates the use of the Lee-Low-Pines transformation~\cite{LLP} in coordinate space: $\hat{\mathcal{H}}'_{\text{pol}}\to e^{i\bm{Q}\bm{r}_0}\hat{\mathcal{H}}_{\text{pol}}e^{-i\bm{Q}\bm{r}_0}$, which removes the coordinate of the impurity from the Hamiltonian. 

For the ground-state manifold ($\bm{Q}=0$), we derive 
\begin{equation}
\hat{\mathcal{H}}'_{\text{pol}}=-\sum\limits_{i=1}^{N}\frac{\partial^2}{\partial \bm{z}_i^2}-\frac{1}{2M}\Bigg(\sum\limits_{i=1}^{N}\frac{\partial}{\partial \bm{z}_i}\Bigg)^2+\sum\limits_{i=1}^{N}U(\bm{z}_i)\,.
\label{eq_H_after_LLP}
\end{equation}
This Hamiltonian describes a complicated many-body problem where the particle-particle interactions are hidden in the mixed derivatives.  To solve the problem, we adopt the following strategy: we assume that $M$ is large, and write $\hat{\mathcal H}'_{\text{pol}}=\hat{\mathcal H}_{0} + \hat{\mathcal H}_P$, where the leading part of the Hamiltonian is
\begin{equation}
\hat{\mathcal H}_{0}=-\frac{1}{2\mu}\sum\limits_{i=1}^{N}\frac{\partial^2}{\partial \bm{z}_i^2}+\sum\limits_{i=1}^{N}U(\bm{z}_i)\,,
\label{eq_ham_main}
\end{equation}
with the reduced mass $\mu\equiv M/(1+2M)$. The perturbative term reads as
\begin{equation}
\hat{\mathcal H}_P=-\frac{1}{2M}\sum\limits_{i,j=1}^N\frac{\partial}{\partial \bm{z}_i}\frac{\partial}{\partial \bm{z}_j}.
\end{equation}
As shown in the Appendix, this approach is related to the mean-field approximation.

The Hamiltonian  $\hat{\mathcal H}_{0}$ describes fermions with the mass $\mu$ interacting with a heavy impurity.  Using results of Sec.~\ref{sec_energy}, we calculate the contribution of $\hat{\mathcal H}_{0}$ to the energy of the Fermi polaron 
\begin{equation}
	\varepsilon_{\text{pol},0}\equiv \expval{\hat{\mathcal H}_{0}-\hat{\mathcal H}_{0}[U= 0]}=\frac{m}{\mu}\varepsilon_{I,\text{res}}\,.
	\label{eq_polaron_energy}
\end{equation}
The expectation value here is defined as
\begin{equation}
	\expval{\hat{O}}\equiv\int\Phi^{*}(\bm{z}_1,...,\bm{z}_N)\hat{O}\Phi(\bm{z}_1,...,\bm{z}_N)\,\mathrm{d}\bm{z}_1,...,\mathrm{d}\bm{z}_N\,,
\end{equation}
where the function $\Phi$ describes the ground state of the Hamiltonian $\hat{\mathcal H}_{0}$.
The contribution from $\hat{\mathcal H}_{P}$ can be estimated within first-order perturbation theory as follows
\begin{equation}
	\varepsilon_{\text{pol},P} = \varepsilon_{\text{pol},0}+\expval{\hat{\mathcal H}_{P}-\hat{\mathcal H}_{P}[U=0]}\,.
	\label{eq_energy_pert}
\end{equation}
This expression has a closed analytical form, which allows for a straightforward evaluation of the self-energy.  

%%%%%%%%%%%%%%%%%%%%%%%%%%%%%%%%%%
\begin{figure}
	\centering
	\includegraphics[width=\linewidth]{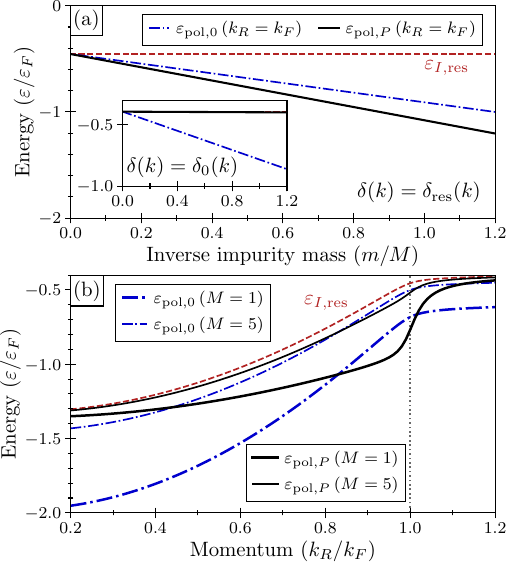}
	\caption{(a)~Dependence of the energy $\varepsilon_{\text{pol}}$ on the inverse impurity mass $M^{-1}$.  We assume that $k_R=k_F$. In the inset, we plot the self-energy of the impurity without a resonant level, see the phase shift in Eq.~\eqref{eq_phase_shift_simple}. (b) Dependence of the self-energy $\varepsilon_{\text{pol}}$ on the position of the resonance $k_R$ for small ($M=1$, thick curves) and large ($M=5$, thin curves) masses of the impurity. 
	In both panels, $\varepsilon_{\text{pol},P}$ are shown as solid black curves,  $\varepsilon_{\text{pol,0}}$ are presented as dash-dotted blue curves, and the self-energies of an infinitely heavy impurity are shown with dashed red curves.
	We use $k_F\alpha=-5$ and $\Gamma=0.05\,\varepsilon_{F}$.}
	\label{fig_mean_field}
\end{figure}
%%%%%%%%%%%%%%%%%%%%%%%%%%%%%%%%

In Fig.~\ref{fig_mean_field}(a), we compare the self-energies with and without the perturbative correction (i.e., $\varepsilon_{\text{pol},P}$ and $\varepsilon_{\text{pol},0}$) to the self-energy of a heavy impurity $\varepsilon_{I,\text{res}}$. We consider the resonant phase shift from Eq.~\eqref{eq_phase_shift} with $k_R=k_F$. Akin to analytical results of Ref.~\cite{Chevy_Fermi_polaron}, the self-energies are inversely proportional to the impurity mass $M$. Note however that without a resonance the dependence on the mass is very weak. The self-energy $\varepsilon_{\text{pol},P}$ is almost identical to the energy of a heavy impurity, see the inset of Fig.~\ref{fig_mean_field}(a). This observation is in agreement with numerical results of Ref.~\cite{Chevy_Fermi_polaron}. A strong dependence on the mass is a direct consequence of the presence of a resonance. In Fig.~\ref{fig_mean_field}(b), we illustrate this further by plotting the dependence of the polaron energy $\varepsilon_{\text{pol}}$ on the resonance momentum $k_R$. We see a strong effect of the mass of the impurity only if the resonance is in the vicinity of the Fermi energy.

\section {Conclusions and Outlook}
\label{sec_conclusion}

We considered a three-dimensional fermionic bath in the presence of a static impurity with a resonance at the Fermi energy. Assuming the Breit-Wigner form of the resonance, we computed several properties of the system. First, we calculated its self-energy using Fumi's theorem. The self-energy is non-analytic in the width of the resonance, which allowed us to suggest emergence of a length scale that determines the physics at intermediate ranges. To investigate this length scale further, we calculated the density of the Fermi gas. The oscillations of the density exhibit universal long-range physics beyond the standard Friedel oscillations, which can be intuitively explained in terms of a tunneling of a fermion to the internal state of the impurity. We argued that the density oscillations can be observed in current cold-atom experiments provided that their temperatures are a fraction of the Fermi temperature. 

Our findings pave a way for a number of future studies. First of all, the resonance should introduce an additional time scale. We did not observe this scale in the quench dynamics of the density, but it must be important in the long-time dynamics of other observables. For example, the Anderson orthogonality catastrophe~\cite{Anderson1967} manifests itself as a decay of a time-dependent overlap~\cite{Nozieres1969,Rivier1971}, see Ref.~\cite{Heavy} for review. The rate of this decay is usually given by the value of the phase shift at the Fermi energy. Fast change of the phase shift due to a resonance at the Fermi surface will introduce a new time scale into this problem. The importance of this scale should be investigated in a future work. It can be even first addressed experimentally, for example, by studying real-time evolution of Ramsey response. The corresponding short-time behavior will be determined only by few-body physics~\cite{Skou2021,Fermi_polaron_experiment_3} but long-time dynamics will reveal the relevant many-body physics.

Second, the density of the Fermi gas dictates the shape of the induced impurity-impurity interactions, at least for weakly interacting heavy impurities. Our study suggests a possibility to strongly modify the density of the bath using the internal structure of the impurity. This may extend the family of existing impurity-impurity interactions mediated by Fermi gases~\cite{Fuchs2007,RKKY_Bose_1,Pasek2019,Huber2019,Enss2020Fermi,Kwasniok2020,RKKY_Bose_2, Edri2020}. Although, our paper focuses on a three-dimensional system, it makes sense to study induced correlations in low-dimensional geometries where the effects of interactions are usually more pronounced. Even an experimental observation of the density oscillations will be easier in one dimension where their decay is slow in comparison to higher spatial dimensions.   

\section*{Acknowledgements}
M.L.~acknowledges support by the Austrian Science Fund (FWF), under project No.~P29902-N27, and by the European Research Council (ERC) Starting Grant No.~801770 (ANGULON). A.G.V.~acknowledges support by European Union's Horizon 2020 research and innovation programme under the Marie Sk\l{}odowska-Curie Grant Agreement No. 754411.\\

\newpage

\appendix

\section{A mean-field approach to the Fermi polaron.}
\label{app}

Here, we provide additional insight into Eq.~(\ref{eq_H_after_LLP}).
To this end, we consider a variational wave function for $\hat{\mathcal{H}}_{\text{pol}}$ in the form of the Slater determinant~\cite{Edwards_MF}:
\begin{equation}
    F(\bm{r}_0,...,\bm{r}_N)=\frac{\exp[i\bm{Q}\cdot\bm{r}_0]}{\sqrt{N!}}
    \begin{vmatrix}
    f_1(\bm{z}_1) & ... & f_1(\bm{z}_N) \\
    \vdots & \ddots& \vdots\\
    f_N(\bm{z}_N) & ... & f_N(\bm{z}_N) 
    \end{vmatrix}\,.
    \label{eq_Slater}
\end{equation}
$\bm{Q}$ is the momentum of the impurity, and $f_i$ are functions that should be obtained variationally. The pre-factor $\exp[i\bm{Q}\cdot\bm{r}_0]$ together with the coordinate shift $\bm{z}_i\equiv\bm{r}_i-\bm{r}_0$ is motivated by the Lee-Low-Pines transformation~\cite{LLP}. 
Note that this mean-field approximation is similar in spirit to the ansatz proposed by Eugene Gross for the Bose gases~\cite{GROSS_1962}, and which was successfully used for a Bose polaron in one~\cite{Volosniev2017,Mistakidis2019,PANOCHKO_2019,Jager2020,Koutentakis2021} as well as in three spatial dimensions~\cite{Guenther2021,Enss2020,Hryhorchak_2020,Massignan2021}. However, there is an important difference in the resulting calculations. Namely, one cannot neglect the term with mixed derivatives when calculating the energy of the Fermi gas, see the equation below. This difference significantly complicates the analysis of the Fermi polaron using the mean-field approximation, although it can still be analytically solvable for one-dimensional problems with a contact interaction~\cite{Edwards_MF}. 

The mean-field energy of the system is given by the expectation value of the Hamiltonian from Eq.~\eqref{eq_hamiltonian_polaron}
\begin{align}
    &\varepsilon_{\mathrm{MF}}\equiv\int\mathrm{d}\bm{r}_0...\mathrm{d}\bm{r}_N F(\bm{r}_0...\bm{r}_N)\hat{\mathcal{H}}_{\text{pol}}F(\bm{r}_0...\bm{r}_N)=\nonumber\\
    &-\frac{1}{2\mu}\sum\limits_{j=1}^N \expval{\frac{\partial^2}{\partial \bm{z}^2}}_{jj}+\sum\limits_{j=1}^N \expval{U(\bm{z})}_{jj}+\frac{1}{2M}\nonumber\\
    \times\Bigg(\bigg[\bm{Q}+&i\sum\limits_{j=1}^N \expval{\frac{\partial}{\partial \bm{z}}}_{jj}\bigg]^2-
    \sum\limits_{j,k=1}^N \expval{\frac{\partial}{\partial \bm{z}}}_{jk}\expval{\frac{\partial}{\partial \bm{z}}}_{kj} \Bigg)\,,
    \label{eq_Hartree}
\end{align}
where we have defined 
\begin{equation}
    \expval{O(\bm{z})}_{jk}\equiv\int f_j^*(\bm{z})O(\bm{z})f_k(\bm{z})\mathrm{d}\bm{z}\,.
    \label{eq_expval}
\end{equation}

In the main text, we focus on the ground-state properties ($\bm{Q}=0$) and a heavy impurity ($M\gg m$).
In this case, we can find the set $\{f_i\}$ that minimizes the expectation value $\varepsilon_{\mathrm{MF}}$ following an iterative procedure. First, 
we remove the last term in Eq.~\eqref{eq_Hartree}, which is equivalent to minimizing $\hat{\mathcal{H}}_0$ from Eq.~(\ref{eq_ham_main}). Calculation of a correction due to the last term is then equivalent to calculating first-order perturbation to the energy due to $\hat{\mathcal{H}}_P$.  Therefore, our approach to the problem can be seen as a mean-field approximation in the frame co-moving with the impurity, at least for a heavy impurity. 
Note that although we focus here on the ground-state properties in this paper, future studies might employ the same approach to study excited states at $\bm{Q}\neq 0$. 

\bibliography{bibliography}
\end{document}